# Building on Quicksand

Pat Helland
Microsoft Corporation
One Microsoft Way
Redmond, WA 98052 USA
PHelland@Microsoft.com

Dave Campbell
Microsoft Corporation
One Microsoft Way
Redmond, WA 98052 USA
DavidC@Microsoft.com

## ABSTRACT
Reliable systems have always been built out of unreliable components [1]. Early on, the reliable components were small such as mirrored disks or ECC (Error Correcting Codes) in core memory. These systems were designed such that failures of these small components were transparent to the application. Later, the size of the unreliable components grew larger and semantic challenges crept into the application when failures occurred.

Fault tolerant algorithms comprise a set of idempotent sub-algorithms. Between these idempotent sub-algorithms, state is sent across the failure boundaries of the unreliable components. The failure of an unreliable component can then be tolerated as a takeover by a backup, which uses the last known state and drives forward with a retry of the idempotent sub-algorithm. Classically, this has been done in a linear fashion (i.e. one step at a time).

As the granularity of the unreliable component grows (from a mirrored disk to a system to a data center), the latency to communicate with a backup becomes unpalatable. This leads to a more relaxed model for fault tolerance. The primary system will acknowledge the work request and its actions *without* waiting to ensure that the backup is notified of the work. This improves the responsiveness of the system because the user is not delayed behind a slow interaction with the backup.

There are two implications of asynchronous state capture:
1) <u>Everything promised by the primary is probabilistic</u>. There is always a chance that an untimely failure shortly after the promise results in a backup proceeding without knowledge of the commitment. Hence, nothing is guaranteed!
2) <u>Applications must ensure eventual consistency</u> [20]. Since work may be stuck in the primary after a failure and reappear later, the processing order for work cannot be guaranteed.

Platform designers are struggling to make this easier for their applications. Emerging patterns of eventual consistency and probabilistic execution may soon yield a way for applications to express requirements for a "looser" form of consistency while providing availability in the face of ever larger failures. As we will also point out in this paper, the patterns of probabilistic execution and eventual consistency are applicable to intermittently connected application patterns.

This paper recounts portions of the evolution of these trends, attempts to show the patterns that span these changes, and talks about future directions as we continue to "build on quicksand".

## 1. Introduction
There is an interesting connection between fault tolerance, offlineable systems, and the need for application-based eventual consistency. As we attempt to run our large scale applications spread across many systems, we cannot afford the latency to wait for a backup system to remain in synch with the system actually performing the work. This causes the server systems to look increasingly like offlineable client applications in that they do not know the authoritative truth. In turn, these server-based applications are designed to record their intentions and allow the work to interleave and flow across the replicas. In a properly designed application, this results in system behavior that is acceptable to the business while being resilient to an increasing number of system failures.

This paper starts by examining the concepts of fault tolerance and posits an abstraction for thinking about fault tolerant systems. Next, section 3 examines how fault tolerant systems have historically provided the ability to transparently survive failures without special application consideration by using synchronous checkpointing to send the application state to a backup. In section 4, we begin to examine what happens when we cannot afford the latency associated with the synchronous checkpointing of state to the backup and, instead, allow the checkpointing of state to be asynchronous. Section 5 examines in much more depth the ways in which an application must be modified to be true to its semantics while allowing asynchronous checkpointing of the application state to its backup. Section 6 looks at a couple of example applications which offer correct behavior while allowing delays (i.e. asynchrony) in checkpointing state to the backup. In section 7, we consider the management of resources when the operations may be reordered due to asynchrony. Section 8 examines the relationship between this class of eventual consistency and the CAP (Consistency, Availability, and Partition-tolerance) Theory. Finally, in section 9, we consider some areas for future work.

## 2. An Abstraction for Fault Tolerance
In section 2, we discuss the broad ideas required to build a fault tolerant system. First, we start by describing the external behavior of the systems we are considering. Next, we describe what it can mean for these systems to offer transparent fault tolerance and not require special application consideration to cope with failures. Then, we quickly consider the issues associated with scalability of these systems. Finally, we briefly discuss the role of transactions in the composition of these fault tolerant systems.

## 2.1 Modeling "The System"

In considering interactions with a fault tolerant "system", we want to look at its behavior as a black-box. From the outside, requests are sent into the system for processing. In years past, these requests looked like block mode screen input. Nowadays, they typically take the form of XML, SOAP, and/or other web-style requests.

To be robust, these incoming requests are retried by their source. In classic fashion, a request is issued and if a timer expires, it is reissued. The fault tolerant server system had better make this work idempotent or the retries would occasionally result in duplicative work. In practice, systems evolve to be idempotent as designers either anticipate the problem or make changes to fix it.

To support this need for idempotence, either each request is submitted with a "uniquifier" that ensures the request is unique (and ensures retries will be associated with the original request), OR the service applies some trick to accomplish the same thing. An example trick is the creation of an MD5 hash of the entire incoming request. With extremely high probability, the MD5 hash is one-to-one correlated with a unique incoming request.

So, the fault tolerant system processes a sequence of requests from an external partner. The requests (and their responses) perform some business task or tasks.

## 2.2 Transparent Fault Tolerance

Fault tolerant systems comprise many components and their design goal is to continue functioning when one (or sometimes more than one) component fails. In this discussion, we will not consider Byzantine Failures [6] in which a component may behave erroneously (and in the Byzantine analysis, potentially maliciously). Instead, we assume *fail fast* [8] in which a component is either functioning correctly or simply stops functioning. Fail fast does not address the possibility of components deliberately misbehaving. Also, it leaves vague the question of what happens when a component performs so slowly that it wreaks havoc on the system. For this discussion, we will address some issues present even with the simplifying assumptions of fail fast.

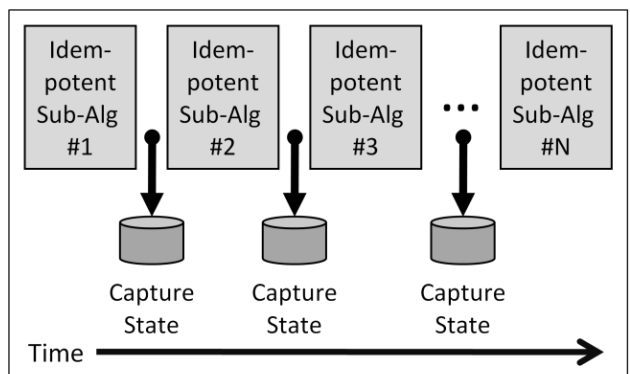

Figure 1) Breaking up a fault tolerant algorithm into sub-algorithms which are each idempotent. By capturing state in between the sub-algorithms and ensuring the state is kept across failures, the larger algorithm can tolerate the faults.

We have observed the pattern in which a fault tolerant algorithm is broken into idempotent sub-algorithms. By capturing sufficient information between the idempotent steps and sending it across the failure boundary, the overarching algorithm can tolerate faults.

From this perspective, you can imagine stepping across a river from rock to rock, always keeping one foot on solid ground. It is important to realize this provides a linear sequence of steps marching forward through the work.

It turns out that many existing fault tolerant systems use this technique to make failures transparent to the application and to the user. We will explore some examples of these systems.

## 2.3 Scaling and Idempotent Sub-Algorithms

In [15] one of the authors (Helland) argues that scalable and distributed applications need special attention when built without distributed transactions. Distributed transactions (especially using the Two Phase Commit protocol [3]) result in fragile systems and reduced availability. For this reason, they are rarely used in production systems, particularly when the resource managers span trust and authority boundaries. In the paper cited above, it is proposed that a scalable application must apply a discipline of partitioning its data into chunks which can remain on a single node even when repartitioned. Each chunk has a unique key.

In the design of fault tolerance systems, we frequently see these idempotent sub-algorithms spread in a distributed fashion around the network. One pattern that has been emerging is that these idempotent sub-algorithms follow the same co-location as described above. All of their data and behavior reside on a single node even in the presence of repartitioning. The collective data will be identified with some unique identifier (called a "key" in [15]) that ensures it remains on exactly one node at a time[1].

## 2.4 Transactions and Idempotence

Transactions can make it EASIER to build idempotent sub-algorithms. Atomic transactions are, well, atomic and do not expose partial results. By using transactions, many (but not all) of the challenges of creating idempotent behavior are eliminated. All that remains is to ensure the work is either not started more than once or that a second attempt will detect a successful first attempt and be innocuous. Some examples of this will be shown.

## 3. Preserving Transparency While Growing

In this section, we examine how the Tandem NonStop system implemented transparent fault tolerance by leveraging synchronous checkpointing across the failure boundaries. We first consider the Tandem system in approximately 1984 when the checkpointing strategy involved sending state to the backup as a part of every individual database WRITE operation. This was correct but had some performance challenges. In roughly 1985, the software was modified to a new strategy which had performance advantages. So, we next examine the behavior of the Tandem systems in approximately 1986 when the checkpointing of state was less aggressive but still sufficient to provide the transparent guarantees. We conclude section 3 with a discussion of why the change from the 1984 to 1986 versions was an erosion of the semantics of failures but was an acceptable erosion of the semantics experienced in a failure.

---

[1] More precisely, the data resides on exactly one node when we ignore the need to do replication underneath the partition… more on that later.

## 3.1 Example 1: Tandem NonStop circa 1984

The Tandem NonStop System is a shared-nothing multi-processor with a message-based interconnect [5]. Each processor has its own CPU, memory, access to the messaging busses, and access to IO-Controllers. Each IO-controllers is dual ported and can be accessed by either of two processors in the system. Pairs of IO-Controllers accessed mirrored disks. This hardware architecture, combined with the Guardian operating system, Transaction Monitoring Facility (TMF), and Disk Process (DP) offered industry leading availability [8] for OLTP systems through the 1980s and continuing today.

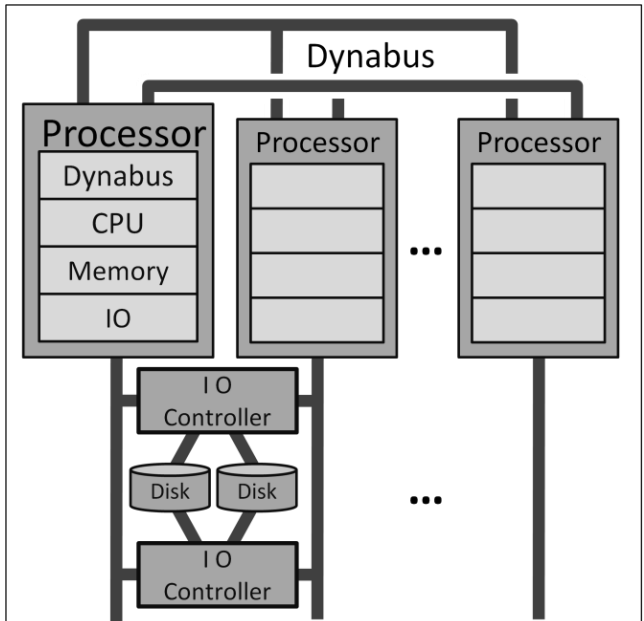

Figure 2) The Tandem NonStop hardware architecture. 2 to 16 processors (with memory and IO-port) are connected via a dual ported message bus called the Dynabus. IO-controllers connect to two processors each. Mirrored disks are connected to redundant IO-controllers. This tolerates any single point of failure.

To perform transactional application work, an app runs in one of the processes and uses messages to do READS and WRITES to the Disk Processes which manage the data and generate log records for the transactional log. Work is actively checkpointed[2] for each WRITE to ensure the backup is able to continue in the event of a failure of the primary disk processor. At transaction commit, all dirtied DPs are asked to flush their log to a centralized ADP (Audit Disk Process).

---

[2] A checkpoint is a technique used to manage state in a Process Pair as described in [12]. Two identical processes run on different processors, one the primary and one the backup. A checkpoint is a message from the primary to the backup describing needed state to ensure fault tolerant service.

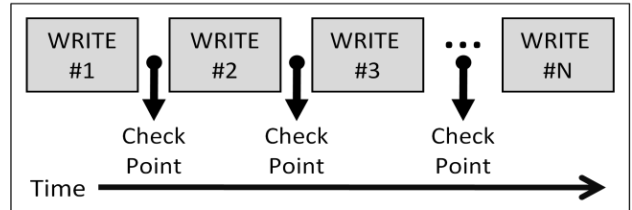

Figure 3) Tandem, circa 1984, each WRITE to the DP is an idempotent sub-algorithm. It is sent across the fallible component (a processor) via a checkpoint from the primary to the backup Disk Process (DP).

It is interesting to note the granularity of the pieces in the approach to fault tolerance. Each WRTE operation is idempotent and, circa 1984, was actively checkpointed to the backup DP [4]. The granularity of the failure is a single process or processor. The granularity of the "idempotent sub-algorithm" is a single WRITE to the DP which gets checkpointed. Failures of a primary DP do not necessarily cause a failure of the transaction.

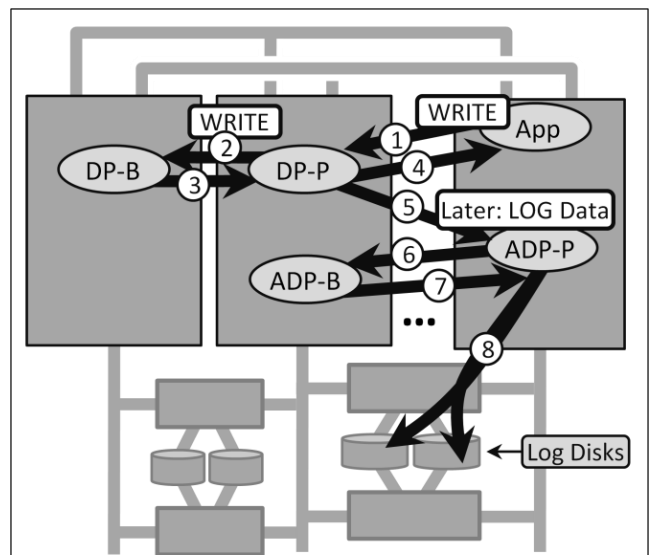

Figure 4) Tandem's system (circa 1984) showing an application process, the Disk Process (DP) primary and secondary, and the Audit Disk Process (ADP) which writes the log to disk. The checkpoint between the primary and backup DPs captures the WRITE in a fashion only loosely correlated to the transaction log.

## 3.2 Example 2: Tandem NonStop circa 1986

In 1985, a new software release of the Tandem NonStop Guardian operating system included a new Disk Process called DP2. This release offered a number of changes including a dramatic optimization in the strategy for fault tolerance [7].

DP2, a completely redesigned disk process, had a whole new approach to checkpointing. The transaction log, describing the changes to the state on disk, was also used to describe the changes that should be known to the backup disk process. In other words, checkpointing and transaction logging were combined into one mechanism. The log would first go to the backup, then to the ADP which would write it on disk.

A design goal was to allow the changes for a transactional WRITE to lollygag within the transactional log in memory of the primary DP2 Disk Process. The WRITE to the primary DP2 could be answered back to the application. Of course, this left open the challenge of correctness if a failure wipes out the primary DP2 with buffered changes done by uncommitted WRITEs. How can this be correct?

The key to the new approach to correctness in this example was to ensure that any in-flight transactions that used a failed primary DP would have their transactions aborted. Since all committed transactions will have pushed their changes to disk, any loss of the memory state when a primary DP fails will only impact an in-flight transaction. Since the system automatically aborts any relevant in-flight transactions when the primary DP fails, correctness is preserved.

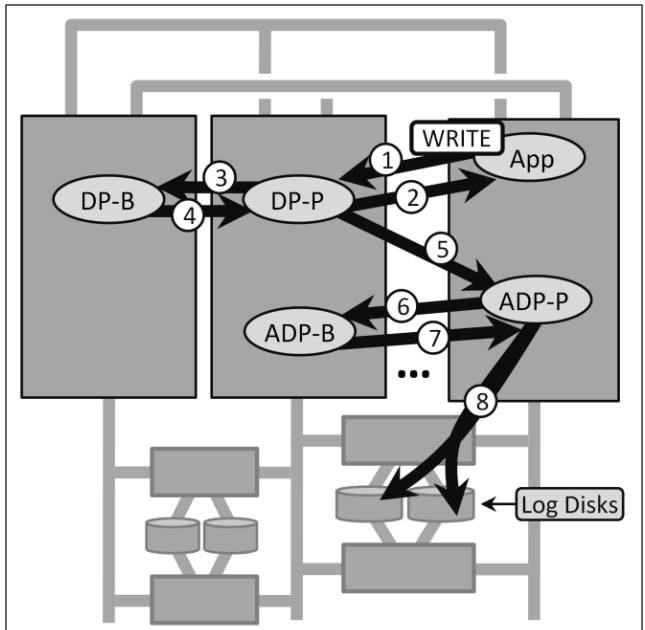

Figure 5) Tandem's system (circa 1986) showing an application process, the Disk Process (DP) primary and secondary, and the Audit Disk Process (ADP) which writes the log to disk. DP checkpoints are the same as log contents. They flow from the primary DP to its backup, to the ADP (primary and backup) and then to disk. Note that the App gets an acknowledgement (2) before all this stuff is sent everywhere it needs to go (even the checkpoint to the backup DP is asynchronous to the response to the App's WRITE request).

This scheme meant that a processor failure would result in more transactions aborting. This was a very rare event and was completely within the system rules which allowed transactions to abort without cause. Arguably, this change was almost transparent to application developers and to users.

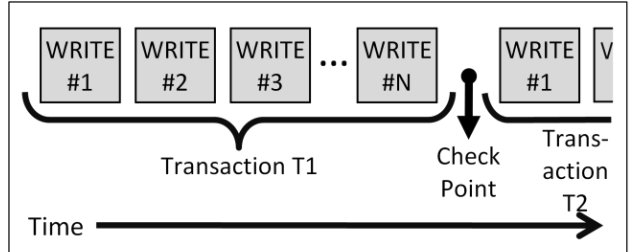

Figure 6) Tandem, circa 1986, each WRITE to the DP is buffered locally within the primary DP. It is NOT guaranteed to be sent to the backup and the loss of the primary results in the transaction's abort. The log for the transaction is guaranteed to be checkpointed as a part of transaction commit. The WRITE is no longer the indempotent sub-algorithm, the transaction is now the idempotent sub-algorithm.

The new scheme offered a huge performance improvement. A WRITE to DP2 could be performed without checkpointing to the backup. This was a dramatic savings in CPU cost and an even more dramatic savings in latency since the application did not need to wait for the checkpoint to see the response to the WRITE. The buffer containing the log entries shifted to being pushed to the backup (and, indeed, to the ADP) on a periodic basis. This is much like group commit [11]. It is easy to understand the efficiency when you think of the difference between a car per driver racing across town versus a city bus sweeping up all the passengers every five minutes or so. As described in [11], waiting to participate in shared buffer writes can, under the right circumstances, result in a reduction of latency since the overall system work is reduced. This reduction in work may reduce system utilization and may more than compensate for the delay.

Looking back at our abstraction for fault tolerance, we see that the idempotent sub-algorithm has grown from a WRITE to a transaction. The granule of failure is still a processor and the application and user barely perceive the change in algorithm.

### 3.3 An Acceptable Erosion of Behavior

Tandem's system in 1986 offered significantly better performance than its predecessor in 1984. Still, there were failures of processors that would yield different behavior than the previous release. When a processor failed in the middle of a transaction, the earlier release would continue forward. Circa 1986, a processor failure may result in the loss of the ongoing transaction.

While this was technically a change in behavior, there were reasons why this was acceptable. All along, the system rules for transactions allowed the transaction to be aborted without (apparent) cause. Deadlocks, operator decisions, timeouts, and other reasons could cause transactions to fail. Because of this, the change was an acceptable erosion of behavior.

The 1984 version of the Tandem system implemented synchronous WRITES to the backup. The WRITE from the user's application was not acknowledged until checkpointed. Circa 1984, the WRITES were asynchronous but the <u>transaction commit</u> was guaranteed to be synchronously checkpointed across the failure boundaries.

## 4. The Creeping Arrival of Asynchrony

In this section, we see the first example of acknowledging the incoming request BEFORE ensuring the work is sent to the backup. This is asynchronous checkpointing to the backup.

We start with a very simple discussion of "log-shipping" wherein the transaction log is sent to a backup system sometime after the user request is acknowledged. This is a fundamental change which deeply impacts the guarantees made to the user.

After introducing log-shipping, we discuss the takeover semantics of this approach. Following this, we look at how this asynchrony means we have to revisit our abstraction for fault tolerance.

### 4.1 Example 3: Log Shipping

This example is well known to most readers. A classic database system has a process that reads the log and ships it to a backup data-center. The normal implementation of this mechanism commits transactions at the primary system (acknowledging the user's commit request) and asynchronously ships the log. The backup database replays the log, constantly playing catch-up.

Typically, applications and users are oblivious to log-shipping. Unless a fault occurs, the application and the user are fat, dumb, and happy. When a fault DOES occur, some recent transactions are lost as the backup takes-over and provides service.

This means that the abstraction described above in section 2 works here except the state is not immediately captured by the backup. Fault tolerance is NOT transparent. It is OK (with low probability) to completely discard recent work.

To allow a datacenter failure to be transparent, the log shipping algorithm would need to stall the response to the commit request at the primary until the primary knows the backup has received the log. This delay is unacceptable in most installations and they deal with the low probability chance of losing recent work. The change from a synchronous transfer of state to an asynchronous transfer is an interesting erosion of the basic abstraction and is another example of where the cost for "consistency at a distance" is too high just as it was when tried to stretch 2PC beyond resource managers in the same room.

> Log-shipping: Our first example where giving a little bit in consistency yields a lot of resilience and scale!

### 4.2 Log Shipping and Takeover Semantics

Log shipping is asynchronous to the response to the client. This inherently opens up a window in which the work is acknowledged to the client but it has not yet been shipped to the backup. A failure of the primary during this window will lock the work inside the primary for an unknown period of time. The backup will move ahead without knowledge of the locked up work.

In most deployments of log-shipping, this is not considered in the application design. It is assumed that this window is rare and that it is unnecessary to plan for it. Bad stuff just happens if you get unlucky. Unfortunately, in most of these systems, a takeover either requires manual cleanup of the work not transferred from the primary to the backup or the work is simply lost.

### 4.3 Revisiting the Abstraction

So, we've seen a basic model for fault tolerance and how it can be applied in a few different systems. In the first two examples, the fallible component was a processor running in the same box as its partner. The close proximity of the components allowed for practical use of synchronous state copying. In the log shipping example, the delay is considered impractical and the transfer of the state is asynchronous. This results in "faults" in the fault tolerance provided for data center failures.

## 5. Loosening the Abstraction

OK… So, maybe there's a broader abstraction at play!

The old model assumed the work would be processed in exactly one order of execution. There was a default "single system of record" form of isolation provided by the classic database system running at the primary. This single history allows for a low-level READ and WRITE semantic that depends on "replaying history".

In this new world, history cannot be exactly replayed and we must count on the ability to reorder the work. This means that we cannot completely know the accurate state of the system. It also means we must move the correctness and reordering semantics up from being based on system properties (i.e. READ and WRITE) to application based business operations.

Section 5 examines a number of different aspects of asynchronous checkpointing and how it impacts application design. We will first discuss the impact of asynchrony on our ability to know the actual true state of the application. Then, we look at probabilistic business rules and how asynchronous checkpointing means we cannot have definitive enforcement of business rules. We discuss the impact of commutativity on the business rules. Next, we consider partitioning of the work and idempotent operations across the partitioned state. After this, we examine the possibility of having a choice of sometimes performing synchronous checkpointing of state if the risk for the specific operation is too great. Next, we consider how the system may handle unanticipated problems and when human intervention may be required. After that, there is a discussion of how asynchronous enforcement of business rules sometimes results in apologies. Finally, we summarize the abstraction by observing that either you have synchronous checkpoints to your backup or you must sometimes apologize for your behavior…

### 5.1 Asynchrony and the Truth

Let's consider those orphaned transactions dawdling in the belly of the failed system in the log-shipping example. They are most certainly out of the picture while the dethroned system (or data center) is unavailable. When it does return, the goal of any recovery policy would be to examine the work in the tail of the log and determine what the heck to do! The backup system has continued and there may be challenges when resurrecting the languishing work. The only way this work can be kept is to ensure that the out-of-order retrying of the work does not break things. In some cases, the pending work is simply discarded due to lack of designed mechanisms to reclaim it! This is part of the implicit consistency model for log-shipping without recovery of lost work.

In the log-shipping example, we see rare cases of work reordering as it is temporarily lost and then resurrected. In the more general case, we see independent work performed at disconnected (or slowly connected) sites which may get reordered as it becomes visible to other systems partnering in the work.

> The deeper observation is that two things are coupled:
> 1) The change from synchronous checkpointing to asynchronous to save latency, and
> 2) The loss of a notion of an authoritative truth.

Back when we had a centralized machine with synchronous checkpointing, we knew <u>the</u> one and only one answer at any given point in time. Allowing for work being locked up in an unavailable backup (née primary) means we don't know the truth.

## 5.2 Probabilistic Business Rules

When we have asynchronous checkpointing, we have windows of failure that mean work may be lost or delayed. When a primary fails, there may be work stuck inside the primary that has not yet been sent to the backup. That work is either lost or delayed.

If a primary uses asynchronous checkpointing and applies a business rule on the incoming work, it is necessarily a probabilistic rule. The primary, despite its best intentions cannot know it will be alive to enforce the business rules.

*When the backup system that participates in the enforcement of these business rules is asynchronously tied to the primary, the enforcement of these rules inevitably becomes probabilistic!*

It is the cost/benefit analysis that values lower latency from asynchronous checkpointing higher than the risk of slippage in the business rule that generalizes the fault tolerant algorithm.

> Distribution + Asynchrony → Probabilities of Enforcement

We are seeing the emergence of applications which take this even farther by increasing disconnection to achieve the economic benefits of loose-coupled parallelism and offline. They are lower latency, more parallel, and more available. They just screw up more often and sort the mess out later. Sometimes, that's good!

## 5.3 Commutativity and Business Rules

In many applications, it is possible to express the business rules of the application and allow different operations to be reordered in their execution. When the operations occur on separate systems and wind their way to each other, it is essential to preserve these business rules.

Example business rules may be: "Don't overbook the airplane by more than 15%" or "Don't overdraw the checking account".

> ***Escrow Locking in Serializable Databases***
>
> Escrow locking is a scheme to increase concurrency while preserving classic transactional ACID behavior.
>
> If you assume a set of commutative operations (such as addition and subtraction), you ensure changes are logged via "operation logging". Operation logging does not capture the before and after images of the value but rather logs "Transaction T1 subtracted $10". If transaction T1 needs to be aborted, the system would add $10 rather than restore the value that existed in the field before T1. In this fashion, the work of multiple transactions can interleave as long as they are doing the commutative operations. If any transaction dares to READ the value, that does not commute, is annoying, and stops other concurrent work.
>
> Escrow locking can be implemented in conjunction with constraints enforcing business rules. Consider addition and subtraction operations with a worst-case minimum and maximum for the value. The system simply needs to track the worst case for all the transactions pending commitment. A new operation will be delayed if it MIGHT cause the value to fall out of bounds with the pending work.
>
> Escrow locking offers crisp semantics because it is functioning on a centralized location and can enforce the worst case outcome of the business rules.

This approach is similar to escrow locking as described in [9]. In escrow locking, commutative operations are allowed as long as they do not violate the constraints of the system. Escrow locking was envisaged as a pessimistic locking scheme which crisply preserved serializable behavior. Escrow locking was implemented in Tandem's NonStop SQL in the late 1980s to support high-throughput addition and subtraction.

> ***WRITES to a database are not commutative!***
>
> The layering of an arbitrary application atop a storage subsystem inhibits reordering. Only when commutative operations are used can we achieve the desired loose coupling. Application operations <u>can</u> be commutative. WRITE is not commutative. Storage (i.e. READ and WRITE) is an annoying abstraction…

## 5.4 Idempotence and Partitioned Workflow

It is essential to ensure that the work of a single operation is idempotent. This is an important design consideration in the creation of an application that can handle asynchrony as it tolerates faults and as it allows loose-coupling for latency, scale, and offline.

Each operation must occur <u>once</u> (i.e. have the business impact of a single execution) even as it is processed (or simply logged) at multiple replicas. One room reservations must (with high probability) result in exactly one room set aside for the guest. One book ordered online should not (very often) result in two books delivered to the customer.

To ensure this, applications typically assign a unique number or ID to the work. This is assigned[3] at the ingress to the system (i.e. whichever replica first handles the work). As the work request rattles around the network, it is easy for a replica to detect that it has already seen that operation and, hence, not do the work twice.

Sometimes, incoming work stimulates other work. For example, processing a purchase order may result in scheduling a shipment. Two replicas may get overly enthusiastic about the incoming purchase order and each schedule a shipment. By uniquely identifying the purchase order at its ingress to the system, the irrational exuberance on the part of the replicas can be identified as the knowledge sloshes through the network. We will see below in the discussions of eventual consistency how this can (probabilistically) be rectified.

---

[3] The experienced reader will realize that this leaves the concern for idempotence in the incoming message from the client as captured at the point of ingress. Retries could cause two or more different replicas to charge ahead to help the user. If the work has no side effects (such as simply reading something), it is OK to do the work multiple times. If the work has side-effects, coordination around a cookie or user-id is usually performed to eliminate duplicates.

To avoid duplicate processing, the uniquifier for the request should be functionally dependent only on the request as seen by the server system. This is possible if the unique id is generated outside the server (e.g. a check number as discussed in Section 6.2) and it is also possible if the server calculates it in a predictable way as discussed in Section 2.1).

The unique identifier of the work (the "uniquifier") has two very important roles:

1) The uniquifier provides the key for partitioning the work in a scalable system.
2) The uniquifier allows the system to recognize multiple executions of the same request. In this fashion, they can be collapsed and the work becomes idempotent.

## 5.5 What's Your Stomach for Risk?

In all these cases, we started with the assumption that the cost to know the truth is prohibitive for the application in question. Hence, we are designing the system and, especially, the application running on the system to probably deliver excellent service and, occasionally, to violate the business rules of the application.

Note that that it is possible to have multiple business rules with different guarantees. Some operations can choose classic consistency over availability (i.e. they will slow down, eat the latency, and make darn sure before promising). Other operations can be more cavalier. Some examples:

- Locally clear a check if the face value is less than $10,000. If it exceeds $10,000, double check with all the replicas to make sure it clears.
- Schedule the shipment of a "Harry Potter" book based on a local opinion of the inventory. In contrast, the one and only one Gutenberg bible requires strict coordination!

> The major point is that availability (and its cousins offline and latency-reduction) may be traded off with classic notions of consistency. This tradeoff may frequently be applied across many different aspects at many levels of granularity within a single application.

## 5.6 Fussing and Whining (but Not Too Often)

So, what the heck does the application DO when its business rules are violated? The application will usually be managing the probabilities so that this is unlikely (since there is frequently a business cost associated with screwing up). Still, this will happen!

The best model for coping with violations of the business rule is:

1. Send the problem to a human (via email or something else),
2. If that's too expensive, write some business specific software to reduce the probability that a human needs to be involved.

While this may sound too simplistic, it is what applications typically do when dealing with these complex issues. It also points out that, in the absence of generalizations of the business rules, the patterns used by the business operations for commutativity, and the business complications of overzealousness, it is not possible to speak to the business consequences or actions to compensate when your luck is bad.

## 5.7 Memories, Guesses, and Apologies

Arguably, all computing really falls into three categories: memories, guesses, and apologies[16, 19]. The idea is that everything is done locally with a subset of the global knowledge. You know what you know when an action is performed. Since you have only a subset of the knowledge, your actions are really only guesses. When your knowledge as a replica increases, you may have an "_Oh, crap_!" moment. Reconciling your actions (as a replica) with the actions of an evil-twin of yours may result in recognition that there's a mess to clean up. That may involve apologizing for your behavior (or the behavior of a replica).

So, consider these three aspects:

- **Memories**: Your local replica has seen what it has seen and (hopefully) remembers it. The cost of spreading that knowledge includes bandwidth, computation, and latency (in the case where you are waiting for the backup to acknowledge your memory of an operation).
- **Guesses**: Any time an application takes an action based upon local information, it may be wrong. This occurs in log-shipping systems where the action is logged locally and has only a very high probability of getting to the backup system before a crash and take-over. That makes it a good guess but it doesn't make it a sure bet. In any system which allows a degradation of the absolute truth, any action is, at best, a guess. It is simply a matter of business choice as to the quality of the guess.
- **Apologies**: When a mistake is made (either due to replication anomalies or because the FAA grounds your jets and you cannot honor your flight reservations), you apologize. Every business includes apologies. As mentioned above, these may be manual with the software enqueuing the problem for human work. Alternatively, application code may issue some apologies for which it has specially designed apology code while asking for human help for those apologies beyond its designed cases.

In a loosely coupled world choosing some level of availability over consistency, it is best to think of all computing as memories, guesses, and apologies.

## 5.8 Synchronous Checkpoints OR Apologies!

So, section 5 is pointing out that there are design options:

1) You can synchronously checkpoint and incur the latency, or
2) You can asynchronously checkpoint, save the latency, and experience modified application semantics.

These modified semantics mean that you don't always know the precise truth because work can be trapped in a partner. They mean you may have to understand that the business rules are enforced probabilistically and may experience reordering. Still, you have the option of sometimes applying business criterion (e.g. the check being for more than $10,000) which cause synchronous checkpointing. Also, it is completely viable to allow human intervention in the resolution of some problems if the chance of this occurring is low enough for this to be cost effective.

In summary, all of these choices depend on their business value!

## 6. Zen and the Art of Eventual Consistency[4]

This section looks in more depth at eventual consistency and how loosely-coupled applications are built to support this. We start with a couple of examples: Amazon's Dynamo storage with a Shopping Cart application on top of it and a classic bank check clearing application. We then talk about applications implementing eventual consistency and contrast this with the difficulties of building eventual consistency in a storage layer. Finally, we discuss the importance of an "operation-centric" approach to eventual consistency in which the operations desired by the user of the application are recorded and become the foundation for the implementation of eventual consistency.

---

[4] We recommend the classic book "Zen and the Art of Motorcycle Maintenance" [2]. A looser, more Zen-like, perspective can be helpful in computing and in one's personal life…

## 6.1 Example 4: Shopping Cart and Dynamo

The Dynamo Storage system [18] is used to support the shopping cart store as well as other systems within Amazon. Dynamo is a replicated blob store implemented with a Dynamic Hash Table (DHT). Dynamo is interesting in many ways including its conscious choice to support availability over consistency. Dynamo always accepts a PUT to the store even if this may result in an inconsistent GET later on.

In [18], the interaction from the application to Dynamo is described as a PUT and GET interface. Due to replication anomalies, a GET may return old information. Performing a PUT based on the old information results in parallel versions. A later GET of this blob may return two sibling (or cousin) versions which the shopping cart application must reconcile. Dynamo, acting as a storage substrate, may present two or more old versions in response to a GET. A subsequent PUT must include a blob that integrates and reconciles all the presented versions.

To do the application level integration, the shopping cart application must record its operations much like a ledger entry. A deletion of an item from the shopping cart is recorded as an operation appended to the cart. These "ADD-TO-CART", "CHANGE-NUMBER", and "DELETE-FROM-CART" operations can usually be reconciled when a union of the operations is finally joined together. Very rare anomalies in the shopping cart are acceptable since the shopper's order is verified as a part of order submission. Unavailability of the shopping cart service is very expensive for Amazon since it results in a drop in business and an unsatisfying user experience.

Dynamo is a storage substrate independent of the shopping cart application layered on top of it. Dynamo returns a blob (and sometimes two or more blobs) to a GET. When more than one is returned, the shopping cart application has to reconcile the confusion AND fold in the new operation it was planning for the cart. The shopping cart application can do this by understanding the contents of the cart as a set of operations. Uniquely referenced operations on the items can be unioned together into a list with a predictable outcome. This is key to the commutativity of operations on the shopping cart. This, in turn, is used to provide exceptionally high availability.

## 6.2 Example 5: Bank Accounts and Ledgers

There is a reason for check-numbers on checks. The check numbers (combined with the bank-id and account-number) provide a unique identifier. Excepting big mistakes (and/or fraud), the payee and amount for a specific check are immutable. The check enters the banking system with a unique identifier and the participants in the loosely-coupled process share information in what may be considered an ongoing workflow.[5]

You deposit your brother-in-law's check for $100 into your bank account and, since you've been a good customer, there is no hold on the money. Your account's balance bumps up from $1000 to $1100. Your bank account information is associated with the check. The check is forwarded to you brother-in-law's bank. Later, when the check bounces, your account is debited $130 (the original $100 plus $30 bounce fee). Interestingly, the decision to be optimistic is based on YOUR good standing with the bank. A less desirable customer (like your brother-in-law) would have a hold placed on the money (reserving for a potential bounce)[6].

Debits and credits to bank accounts are commutative. There is an expressed business rule that the account balance will not drop below zero. If you had spent the $1100 before your brother-in-law's check was returned, your account would have violated the business rule when your brother-in-law's check bounced. Banking policies make this less likely but not impossible. It is a business decision on the part of the bank to allow this risk.

Consider, too, the ledger associated with a bank account. At the end of the month, a statement is issued. It is not critical that it be perfect. Some check floating on midnight of the $31^{st}$ of the month may land in this month's statement or in next month's statement.

The monthly statement starts out with a balance. Debits and credits are applied. Once it is issued, it is permanent and immutable. Errors in March's statement may be adjusted in April's statement but March's statement is never modified.

The bank has two jobs to do with the account. First, it needs to decide if a check should clear based upon the best knowledge of the account's balance. Second, it needs to meticulously remember all the operations (debits and credits) performed on the account.

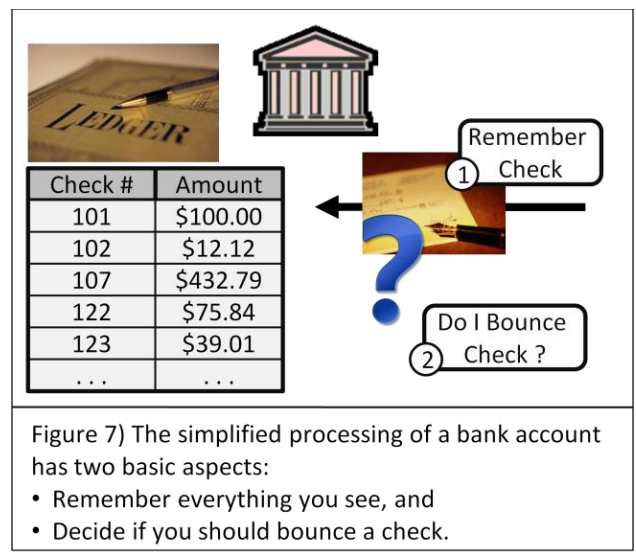

Figure 7) The simplified processing of a bank account has two basic aspects:
- Remember everything you see, and
- Decide if you should bounce a check.

Imagine a replicated bank system which has two (or more) copies of my bank account, both of which are clearing checks. There is a small (but present) possibility that multiple checks presented to different replicas will cause an overdraft that is not detected in time to bounce one of the checks. Each replica that clears a check will remember the check with its check number. Assuming no replica is permanently destroyed, the information about the check will be added to the bank statement and funding allocated for it. A very untimely outage could result in the check landing in next month's statement rather than this month but that's no big deal.

---

[5] This mechanism has been used for many years and pre-dates computerized systems. It was used by our grandparents for the same reason that we advocate the use of a unique-id that is functionally dependent on the incoming request (either by being part of the request or by being derived from the request). The check-number (along with bank and account number) is a wonderful unique-id.

[6] It does make sense to base the decision on YOUR standing… It's the only information available locally to the bank and YOU are more likely to eat the cost than the bank.

In this banking system, the information about the checks can be coalesced as the replicas communicate. The usage of check numbers makes the processing of the check idempotent. The nature of the operations (i.e. addition and subtraction) ensures the work is both commutative AND associative.

It is VERY likely that a banking system layered on flakey computers without raised floors, operators, or backup power would be more cost effective than a high-priced centralized one. As demonstrated above, the dissection of the work into memories, guesses, and apologies is exactly how banks function today.

### 6.3 Eventual Consistency and Applications

In both the Dynamo/ShoppingCart example and in the banking example, uniquely identified work arrives into the system and is processed by a single replica. The work is propagated to other replicas as connectivity allows.

Because the requests are commutative (i.e. reorderable), it is OK for them to be processed at different replicas in different orders. The chances are very high that the result will be the same. Also, the design of each system includes business policies for resolving what will happen if the different orders of execution result in different answers.

### 6.4 Eventual Consistency and Storage

It is interesting that much of the literature on eventual consistency focuses on READ and WRITE semantics. For example, a recent paper [20] by Werner Vogels explains many great concepts in the area of eventual consistency but still couches the discussion in the context of clients, storage systems, and updates.

The Amazon Dynamo paper [18] covers many fascinating topics and shows how Dynamo is a key-value blob store. Still, section 4.4 on Data Versioning discusses the use of "*add to cart*" and "*delete item from cart*" and how these operations are captured within the blobs being stored by Dynamo. Even if the version histories are reordered, items added to the cart will not be lost once their stored blob version is collapsed together with the other versions. Occasionally deleted items will reappear.

Storage systems alone cannot provide the commutativity we need to create robust systems that function with asynchronous checkpointing. We need the business operations to reorder. Amazon's Dynamo does not do this by itself. The shopping cart application <u>on top of the Dynamo storage system</u> is responsible for the semantics of eventual consistency and commutativity.

The authors think it is time for us to move past the examination of eventual consistency in terms of updates and storage systems. The real action comes when examining application based operation semantics.

### 6.5 The "Operation-Centric" Pattern

Both the Dynamo/Shopping-Cart example and the Banking example show the capture of the application's desires.

The Shopping-Cart example uses "*add-to-cart*" and "*delete-item-from-cart*" operations to capture the intention of the user as they add and delete items from the cart. In the Dynamo blob store, the usage of this *operation-centric* approach offers resiliency to the occasional interleaving of versions that results from replication combined with choosing availability over consistency. It is interesting to note that operation-centric work can be made commutative (with the right operations and the right semantics) where a simple READ/WRITE semantic does not lend itself to commutativity.

In the Banking example, "*debit*" and "*credit*" operations are processed at separate replicas and then shared as soon as possible. Again, this is an operation-centric approach which results in commutativity of the operations. Allowing the loose-coupled processing of the "*debit*" and "*credit*" operations will occasionally (but rarely) result in a cleared check that we hope would have bounced. Still, for many environments, this is an acceptable business expense and may be codified in a business rule for loosely coupled systems.

## 7. Managing Resources with Asynchrony

This section looks at how resources are managed in the face of asynchronous checkpointing and the possibility of independent work and potentially redundant work.

First, we consider how resources get allocated across loosely-coupled systems and whether they are over-booked or over-provisioned. After this, we observe that the computers' opinions do not necessarily map to the real world in the face of accidents and other mistakes. Next, we look at one example pattern of conservative (over-provisioned) resource management in the "seat-reservation" pattern. Then, we consider the advantage of making resources similar (or "fungible") whenever possible. Following this, we look at the use of unique identifiers in the management of asynchronous requests. Moving along, we consider eventual consistence and asynchronous management of resources. Finally, we consider the patterns that existed in business before the rise of computers and their use as inspiration in our design of loosely-coupled systems.

### 7.1 Over-Booking versus Over-Provisioning

As we consider a system with asynchronous checkpointing, we are considering a system with a probability that two or more replicas will be allocating resources to their users. Since these replicas will sometimes be incommunicado, we must consider the policy used for allocating resources while not in communication. There are two approaches:

1) <u>Over-Provisioning</u>. In this approach, each replica has a fixed subset of the resources that it may allocate. If there are 1000 books in inventory, each of the two replicas may have 500 to sell. When over-provisioning, a replica <u>cannot</u> make the mistake of allocating a resource that is not truly available to be allocated. On the other hand, over-provisioning means that there <u>will</u> be excess resources kept within the replicas.

2) <u>Over-Booking</u>. Unlike over-provisioning, over-booking allows for the possibility that the disconnected replicas will occasionally promise something they cannot deliver. By allowing independent allocation without ensuring strict partitioning of the resources, sometimes commitments are made that cannot be kept. On the other hand, sometimes business will be scheduled that would be declined under the strict partitioning of over-provisioning.

It is possible to be conservative and ensure you NEVER have to apologize to your customers. This will, however, sometimes result in you deciding to decline business you would rather have. You may accept the business on a disconnected replica without the confidence that you will be able to keep your commitments. You can dynamically slide between these positions (while you are connected) and adjust the probabilities and possibilities.

In the face of disconnection, you cannot know the perfect answer and must adopt a business policy that allows for the tradeoffs that are right for your business!

CIDR Perspectives 2009

## 7.2 Computing versus Reality

In the previous section, we described how over-provisioning of resources means you cannot make the mistake of allocating a resource that is not truly available. This is true in that the computational resource will not show an allocation for which there are no resources. Unfortunately, the real world is not always accurately modeled in the computers (and cannot always be).

Consider a case where the only book in inventory is scheduled for delivery. Due to an over-provisioning scheme, there is no confusion about the inventory and the book is promised to a customer. In preparing the book for shipment, it is run over by the forklift in the warehouse. So, over-provisioning notwithstanding, you need to apologize!

Even if the computer systems are perfect, business includes apologizing because stuff will go wrong!

## 7.3 The "Seat Reservation" Pattern

In certain real world transactions, the resources involved are not considered fungible. One good example is reserving a seat at a concert where the actual seat(s) are considered critical to the buying decision. This sets up a condition where prime seats are crucial both from the perspective of the supplier and consumer. One way to implement a correct application which maintains the business rule that any seat must either be "available" or "occupied and associated with a valid purchase" is to use a database transaction to scope and protect the business rule. This scheme works if you have a trusted agent handling the purchase. Ticketing agents would control the transaction and reserve potentially available seats. If a purchaser reneges, the transaction is rolled back making the seats available for purchase again.

Online buying situations, where the consumer is in control, changes the transaction in both space and time in ways that break our ticketing system in several ways. First off, in the space dimension, the consumer is not a trusted agent. Our transaction now extends beyond the trust boundary originally established for the system business rule. Secondly, in the time dimension, we have no way to constrain the time in which untrusted agents can hold our system in an inconsistent state. In this particular example, untrusted agents could exploit these aspects of the system to quickly start a set of transactions against prime seats, making them unavailable to others, and then reselling them at a profit. Seats that are not sold by our unscrupulous agent could be released by simply rolling back transactions at no cost.

Anyone who has purchased tickets online will recognize the "Seat Reservation" pattern where you can identify potential seats and then you have a bounded period of time, (typically minutes), to complete the transaction. If the transaction is not successfully concluded within the time period, the seats are once again marked as "available". This is done by using three states for seats:

1. {"available"}
2. {"purchase pending", session-identity}
3. {"purchased", purchaser-identity}

Individual database transactions are used to transition from one state to another and to durably enqueue requests to clean up seats abandoned in the "purchase pending" state.

This is an example of sliding the allocation spectrum towards the "over-provisioning" side. The seats are considered unique and coordination is mandated between the primary and backup ensuring a conservative (i.e. "over-provisioned") management of the resources. To avoid this challenge, we need to make a pool of resources.

## 7.4 The Quest for Fungibility

As you look at functions in the computing world, you see an ever increasing categorization of things into fungible buckets. You can't reserve room 301 at the Hilton but you can get a king sized non-smoking room.

Consider a pork-belly. What the heck is a pork-belly? It is a term to describe a bunch of pork[7]. By standardizing a collection of pork, it is possible to sell a pig before it is grown. The existence of a pork-belly as a unit of trade has offered powerful financial mechanisms in support of the raising and distribution of pigs. Farmers can sell their pigs in advance of their maturation and moderate their risk.

The real world is rife with algorithms for idempotence, commutativity, and associativity. They are part of the lubrication of real world business and of the applications we must support on our fault tolerant platforms. A major trick is to look for mechanisms to create equivalence of the operation or resource.

## 7.5 The Importance of Uniquifiers

One important pattern in the management of asynchrony is the usage of the unique identifier in tracking the request through the distributed system. Sometimes the over-zealous replicas will both do the work for a single request and only later detect that this work has been duplicated. If the work has allocated a fungible resource, the system needs to detect this and return the redundantly allocated resource. The detection of the redundant work is made possible by the uniquifier on the request.

So, even as we look at the topic of managing resources under asynchrony, we see the importance of having uniquely identified requests so we can create idempotent behavior.

## 7.6 Eventually We'll Talk and Be Consistent

As disconnected replicas work independently, they accumulate operations. This is identical to a replica of the bank account clearing some checks but not others. It is also identical to a blob in the Dynamo store having some of the operations to the shopping cart (e.g. "ADD-TO-CART" and "DELETE-FROM-CART") but missing out on other operations due to the timing of management of the replication.

When the work flows together, a new, more accurate answer is created. When an application is built to support eventual consistency, the design should ensure that the order of the work's arrival at the node is not the determining factor in the outcome. Replicas that have seen the same work should see the same result, independent of the order in which the work has arrived.

As mentioned above, sometimes the operations accumulated by different replicas result in a violation of the application's business rules. The bank account with independent replicas clearing checks may find an overdraft on the account that, in a centralized system, would have resulted in a bounced check rather than clearing too many checks for the available funds. This level of violation of the business rules becomes a probabilistic analysis with the application designers choosing their stomach for risk.

---

[7] Wikipedia describes a pork belly as the underside of a pig from which bacon is made. A unit of trade is 20 tons of frozen, trimmed bellies. See [21].

## 7.7 Back to the Future

Whenever the authors struggle with explaining how to implement loosely-coupled solutions, we look to how things were done before computers. In almost every case, we can find inspiration in paper forms, pneumatic tubes, and forms filed in triplicate.

Consider the lost request and its idempotent execution. In the past, a form would have multiple carbon copies with a printed serial number on top of them. When a purchase-order request was submitted, a copy was kept in the file of the submitter and placed in a folder with the expected date of the response. If the form and its work were not completed by the expected date, the submitter would initiate an inquiry and ask to locate the purchase-order form in question. Even if the work was lost, the purchase-order would be resubmitted without modification to ensure a lack of confusion in the processing of the work. You wouldn't change the number of items being ordered as that may cause confusion. The unique serial number on the top would act as a mechanism to ensure the work was not performed twice.

## 8. CAP and ACID2.0

As mentioned above, the CAP Theorem [13, 14] states that with Consistency, Availability, and Partition tolerance you can have any two at once but not three. We do not argue with this.

What is interesting is that the consistency for this is based upon the classic ACID. That is: Atomic, Consistent, Isolated, and Durable. The classic ACID has the goal to make the application perceive that there is exactly one computer and it is doing nothing else while this transaction is being processed.

Consider the new ACID (or ACID2.0). The letters stand for: Associative, Commutative, Idempotent, and Distributed [17]. The goal for ACID2.0 is to succeed if the pieces of the work happen:

- At least once,
- Anywhere in the system,
- In any order.

This defines a new KIND of consistency. The individual steps happen at one or more system. The application is explicitly tolerant of work happening out of order. It is tolerant of the work happening more than once per machine, too.

Notice that examples 4 (the shopping cart) and 5 (banking) meet this style of consistency.

### 8.1 Fault Tolerance on ACID

To maintain serializability, classic algorithms do one thing at a time. All the concurrency mechanisms we know and love work hard to provide an appearance that one thing happens at a time.

When considering the fault tolerant abstraction described in section 2, we see the focus on synchronous checkpointing and a linear history. When correctness is defined by the classic ACID, it is essential to provide an ordering amongst the transactions. This is, of course, serializability [3, 10, 12].

It is interesting to contrast example #1 (Tandem circa 1984) and example #2 (Tandem circa 1986). In example #1, the synchronous checkpointing of state to the backup occurs on a WRITE by WRITE basis. This is correct but not necessarily highly performant. In example #2, the synchronous capture of state is at the completion of the transaction. Intra-transaction concurrency is managed by classic database management techniques. Only when completing a transaction is it necessary ensure the work survives a fault.

### 8.2 Fault Tolerance on ACID2.0

OK, now let's consider what happens to the concept of fault tolerance when you are NOT shooting for serializability (classic ACID) but rather for the new ACID of Associativity, Commutativity, Idempotence, and Distribution.

If you break the algorithm for the desired work into pieces, each piece must be idempotent (just like in the basic approach to fault tolerance). Furthermore, we are considering having the work distributed around the network rather than concentrated within a centralized system. The only way to do this while preserving the old guarantees of classic ACID is with the well documented pessimistic or optimistic concurrency control mechanisms. These tend to be fragile.

When the application is constrained to the additional requirements of commutativity and associativity, the world gets a LOT easier. No longer must the state be checkpointed across failure units in a synchronous fashion. Instead, it is possible to be very lazy about the sharing of information. This opens up offline, slow links, low quality datacenters, and more.

Surprisingly, we find that many common business practices comply with these constraints. Looking at the business operations from the standpoint of how work has traditionally been performed shows many examples supportive of this approach. It appears we in database-land have gotten so attached to our abstractions of READ and WRITE that we forgot to look at what normal people do for inspiration.

## 9. Future Work

It seems that it would be of great value to dissect different applications in business environments to see the recurring patterns. What are the operations in play for various applications? When are they commutative? What practices make the operations idempotent? Are there different solutions that are recast syntactically in different environments? Is there a taxonomy of patterns into which the various solutions can be cast?

Our forefathers were VERY smart and were dealing with loosely coupled systems to implement their businesses. They knit the loosely-coupled systems together with messages, telegrams, letters, and the postal system. To cope, they needed reorderable operations. Sometimes, the work was requested twice and this required protocols to implement idempotence. How were these schemes used to run a railroad and build a Model-T? Are these patterns still there waiting for us to use in our distributed systems?

## 10. Conclusion

We have reviewed some of the evolution in highly available and fault tolerant systems. As far as we can see, all reliable systems are built out of a collection of unreliable components which are stitched together in a fashion that provides service in the face of the failure of some of these unreliable components. Over time, the size of the failure unit has gotten larger and larger.

For years, the state of the art in fault tolerant systems provided crisp transactional behavior by synchronously checkpointing state across the failure boundaries. As the size of the failure unit has increased, the latency involved in synchronous checkpointing has grown to be punitive.

In response to the increased latency, applications have embraced the asynchronous transmission of state across failure boundaries. That has required new models and patterns of application design.

We have attempted to describe the patterns in use by many applications today as they cope with failures in widely distributed systems. It is the reorderability of work and repeatability of work that is essential to allowing successful application execution on top of the chaos of a distributed world in which systems come and go when they feel like it. Application designers instinctively gravitate to a world of eventual consistency (usually without the formalisms to help them get there).

Finally, we have examined this with respect to the CAP theory and described how, in this new world, many solutions are designed to take a relaxation of classic consistency to preserve both availability and partition tolerance. This relaxed notion of consistency is very valuable and deserves more academic work.

## 11. ACKNOWLEDGMENTS

We would like to thank Dexter Barnes and Swami Sivasubramanian for their comments on this paper.